# Coulomb interactions in single, charged self-assembled quantum dots: radiative lifetime and recombination energy

P. A. Dalgarno[1], J. M. Smith[1,2], J. McFarlane[1], B. D. Gerardot[1], K. Karrai[3],
A. Badolato[4], P. M. Petroff[4] and R. J. Warburton[1]

[1]*School of Engineering and Physical Sciences, Heriot-Watt University, Edinburgh, EH14 4AS, UK*
[2]*Department of Materials, University of Oxford, Parks Road, Oxford, OX1 3PH, UK*
[3]*Center for NanoScience, Department für Physik der LMU, Geschwister-Scholl-Platz 1, 80539 Munich, Germany*
[4]*Materials Department, University of California, Santa Barbara, California 93106*

We present results on the charge dependence of the radiative recombination lifetime, $\tau$, and the emission energy of excitons confined to single self-assembled InGaAs quantum dots. There are significant dot-to-dot fluctuations in the lifetimes for a particular emission energy. To reach general conclusions, we present the statistical behavior by analyzing data recorded on a large number of individual quantum dots. Exciton charge is controlled with extremely high fidelity through an n-type field effect structure, providing access to the neutral exciton ($X^0$), the biexciton ($2X^0$) and the positively ($X^{1+}$) and negatively ($X^{1-}$) charged excitons. We find significant differences in the recombination lifetime of each exciton such that, on average, $\tau(X^{1-}) / \tau(X^0) = 1.25$, $\tau(X^{1+}) / \tau(X^0) = 1.58$ and $\tau(2X^0) / \tau(X^0) = 0.65$. We attribute the change in lifetime to significant changes in the single particle hole wave function on charging the dot, an effect more pronounced on charging $X^0$ with a single hole than with a single electron. We verify this interpretation by recasting the experimental data on exciton energies in terms of Coulomb energies. We show directly that the electron-hole Coulomb energy is charge dependent, reducing in value by 5-10% in the presence of an additional electron, and that the electron-electron and hole-hole Coulomb energies are almost equal.

PACS numbers(s): 71.35.Pq, 73.21.La, 73.23.Hk, 78.67.Hc

## Introduction

The potential for an optically active quantum dot to form the functional device in modern technologies such as quantum information processing has been well documented in recent years[1]. Possible applications may utilize the antibunched photons emitted by single excitons[2,3] or entangled photon pairs from the biexciton decay[4,5]. The practicality of such devices relies on a detailed understanding of the exciton dynamics. A key parameter is the exciton radiative recombination lifetime, $\tau$. The recombination lifetimes of neutral excitons, $X^0$, and biexcitons, $2X^0$, have been reported in numerous quantum dot systems[6-10]. However, there is limited data available on the recombination lifetimes of the positively charged and negatively charged excitons, $X^{1-}$ and $X^{1+}$, respectively. Data on these excitons relative to $X^0$ allow the effects of electron and hole charging to be assessed independently. Ensemble spectroscopy is unsuitable to gain this information as it is generally impossible to control the charge on all the dots in the sample. Furthermore, the energetic shifts in the photoluminescence energy on charging are typically much smaller than the width of the ensemble spectrum. Single dot spectroscopy eliminates these problems and allows $\tau(X^{1-})$, $\tau(X^0)$, $\tau(X^{1+})$ and $\tau(2X^0)$ to be measured on the same dot. However, there are large dot-to-dot fluctuations in the lifetimes. In order to obtain a balanced picture of dot behavior on charging, a large number of dots must be studied.

We present here a comprehensive study of the recombination lifetime from $X^0$, $X^{1-}$ and $X^{1+}$ in addition to $2X^0$. We present detailed measurements on almost 80 InGaAs/GaAs dots from two separate samples. For every dot studied, $\tau(X^{1+})$ is found to be significantly larger than $\tau(X^0)$. In turn, $\tau(X^{1-})$ is, on the average, slightly larger than $\tau(X^0)$. These results demonstrate that the single particle hole wave function is significantly perturbed on charging. We verify this interpretation by determining the Coulomb energies from the emission and charging energies. We find that the electron-hole Coulomb energy is decreased slightly by the addition of an additional electron. Furthermore, the electron-electron and hole-hole Coulomb energies are almost equal.

## Dot Details

The dots studied are molecular beam epitaxy-grown InGaAs dots emitting close to 1.3 eV (950 nm). The emission energy is red-shifted from as grown InAs dots to allow acceptable photon detection efficiencies with silicon based detectors. The red-shifting is achieved by capping the dots with 3 nm of GaAs and annealing for 30 seconds at the growth temperature. Dot size is estimated to be approximately 25 nm x 25 nm in the plane and 2 to 3 nm high[11]. The dots are embedded in an n-type field effect structure. A 25 nm tunneling barrier separates the dots from a heavily doped n-type back contact. Either a 10 nm (sample A) or 30 nm (sample B) capping layer then separates the dots from an AlAs/GaAs blocking barrier. A semi-transparent NiCr Schottky gate (3-8 nm thick) is evaporated onto the sample surface. A d.c. bias applied between the Schottky gate and back contact shifts the dot levels with respect to the Fermi level in the back contact. When the dot level is resonant with the Fermi level, electrons are free to tunnel in and out of the dot. The small physical size of the dots results in a pronounced Coulomb blockade, giving rise to single electron control over the charging. Under non-resonant excitation, a positive space charge region in the device is formed in the valence band between the capping layer and the blocking barrier. Previous work has demonstrated that by taking advantage of this space charge region it is possible to form both positively and negatively charged excitons in the same dot[12, 13]. In Sample A hole tunneling from the dots to this space charge region is suppressed and $X^{1+}$ decay is determined by spontaneous recombination[14]. This is not the case in sample B but sample B was found to have a much stronger biexction emission than sample A and was therefore used to access $\tau(2X^0)$. Both sample A and sample B have a bi-modal distribution of dot emission energies extending from 1.26 eV to 1.38 eV. Using a combination of data from both sample A and B, a full statistical picture of the recombination lifetime behavior of $X^0$, $2X^0$, $X^{1-}$ and $X^{1+}$ as a function of dot emission energy is constructed.

### Experimental Details

The photoluminescence (PL) from a single dot was probed using a diffraction-limited confocal microscope. The collection spot size, using an objective lens of numerical aperture 0.65, was measured to be 780 nm at wavelength 950 nm. Isolation of individual dots was achieved through the use of the high spatial resolution and samples with dot densities of less than 1 dot/$\mu m^2$. Solid immersion lens (SIL) technology was utilized to improve both the spatial resolution and the PL collection efficiency. We note that there was no systematic difference in the lifetimes recorded with/without a SIL demonstrating that the abrupt change in refractive index at the semiconductor surface does not play a significant role[15]. PL spectra were recorded using a dispersive 0.5 m spectrometer (spectral resolution 60 $\mu$eV at 1.3 eV) and a liquid nitrogen-cooled camera. All experiments are performed at 4.2 K.

Lifetime dynamics were measured using time correlated single photon counting (TCSPC). The sample was excited non-resonantly with an 826 nm pulsed laser diode with timing jitter of under 100 ps. Photon counting was performed using a commercially available silicon single photon avalanche diode (SPAD) with a dark count rate of under 50 counts/s. A second exit slit on the spectrometer, accessed via a retractable mirror, was used to direct spectrally filtered light with a bandwidth of 0.5 meV (~ 0.4 nm) to the SPAD. The 0.5 meV filter window allowed for the collection of all the light from a single PL emission line, simultaneously rejecting PL from different charge configurations. The SPAD has a timing jitter of 400 ps and determines the overall temporal response shown in Fig. 2. The power density was kept low enough to avoid saturation effects through multi-exciton cascade.

### Experimental Results

Figure 1a shows the time-integrated PL from a single dot from sample A as a function of bias. The spectral shifts in the PL correspond to single electron charging. The exciton lines, $X^0$, $2X^0$, $X^{1-}$ and $X^{1+}$ are labeled. Exciton lines are identified through signatures in the Coulomb blockade[12, 16, 17] and verified through absorption spectroscopy[18], the PL power dependence[19] and dark exciton decay dynamics[20]. We stress that there is no ambiguity in the identification of the PL lines in these experiments.

TCSPC was performed on each exciton line for the dot shown in Fig 1a as a function of bias. For the complete extent of the gate voltage plateau, each exciton shows a bias-independent primary lifetime[14]. Decay via carrier tunneling would have a strong bias dependence allowing us to deduce that the decay is dominated by spontaneous recombination. $X^0$ decay shows in addition a bias-dependent secondary lifetime which is caused by the dark exciton dynamics[20]. $X^{1-}$ and $2X^0$ have no dark states and therefore show no secondary lifetime at any bias. $X^{1+}$ has a secondary lifetime of 6.7 ns, but it is largely insensitive to bias and is most likely caused by hole recapture after the primary recombination event[21]. Normalized TCSPC data are shown in Fig. 2. The raw data show only slight differences in the decays from the 4 excitons owing to the jitter in the SPAD. However, the signal:noise is high enough that convolution fitting provides lifetimes down to a few hundred ps with 5% uncertainty. The fitted decays are shown in Fig. 2 and return lifetimes of 0.79 ns, 0.84 ns, 0.87 ns and 0.58 ns for $X^0$, $X^{1-}$, $X^{1+}$ and $2X^0$, respectively.

We verify that the results shown in Fig 2 are representative by studying ~80 dots from both sample A and B. All data were taken under similar experimental conditions and for both samples the PL lines span the entire ensemble PL. Fig. 3 shows the recombination lifetimes as a function of the $X^0$ PL energy. There is dot-to-dot scatter in the measured lifetimes arising from the inhomogeneous nature of the dots. However, a clear dependence on both dot charge and emission energy is evident.

To clarify the change in lifetime with varying exciton configuration, lifetime ratios are shown in Fig 4. Averaging over all dots, we find $\tau(X^{1-}) / \tau(X^0) = 1.25 \pm 0.18$, Fig 4a. Fig 4b shows the ratio $\tau(X^{1+}) / \tau(X^0)$, which averages $1.58 \pm 0.55$. For every dot studied, $\tau(X^{1+})$ is greater than $\tau(X^{1-})$. This is highlighted in Fig 4d by comparing directly $\tau(X^{1+}) / \tau(X^0)$ with $\tau(X^{1-}) / \tau(X^0)$. Fig 4c shows the ratio $\tau(2X^0) / \tau(X^0)$ which averages $0.65 \pm 0.1$, consistent with previously reported values[8-10] for similar, but non charge tunable, InGaAs dots. The changes in lifetimes are accompanied by changes in emission energy. The shifts in PL energy for $X^{1-}$, $X^{1+}$ and $2X^0$ relative to $X^0$ are shown as a function of $X^0$ PL energy in Fig. 5.

### Confinement Limits

The properties of an exciton in a quantum dot are dominated by confinement. In the limit of strong confinement, the single particle energy is significantly larger than the exciton binding energy. The electron and hole wave functions are determined by the confining potential and only slightly perturbed by the Coulomb interactions[22]. In this limit, the oscillator strength is related to the overlap integral of the electron and hole wavefunctions, $|\langle \Psi_e | \Psi_h \rangle|^2$, by[23]:

$$f_{osc} = \frac{|\langle \Psi_e | \Psi_h \rangle|^2 E_p}{2E_{PL}} \quad (1)$$

where $E_p$ is the Kane energy (25.7 eV for GaAs) and $E_{PL}$ the dot emission energy. In this limit, the oscillator strength, equivalently the recombination lifetime, is independent of exciton charge and the biexciton lifetime is exactly half the exciton lifetime owing to the two decay channels open to the biexciton.

In the other extreme, the limit of weak confinement, the dot size is much larger than the exciton Bohr radius, approximately 13 nm for GaAs. In this case, the exciton binding energy dominates over the single particle energies. The picture is now one of an exciton as a bound composite particle moving freely in a potential landscape determined by the dot. In this limit, the exciton picks up a contribution to its oscillator strength from each unit cell of the dot, a superradiant effect[24,25], resulting in large oscillator strengths, equivalently small recombination lifetimes. In the weak confinement limit, variations in the exciton charge significantly change the electron-hole correlations, leading to a strong dependence of recombination lifetime on charge.

The self-assembled dots studied here are a few tens of nanometers in size and are therefore far from the weak confinement limit[26]. However, the lifetime data allow us to conclude that the dots are not in the strong confinement limit. First, the recombination lifetime is clearly charge dependent, Fig 4, evidence of charge-dependent carrier wave functions. Secondly, measured on ~30 dots, $\tau(2X^0) / \tau(X^0) = 0.65 \pm 0.1$ ns, Fig. 4c, larger than 0.5 by ~1.5 standard deviations. Similarly, other experiments have found ratios of 0.5 to 0.7[6-10], albeit from fewer dots than were measured here. Recent theoretical work has successfully modeled these lifetime ratios within an intermediate confinement regime, using either path-integral Monte Carlo integration[27] or configuration interactions amongst atomistic wavefunctions[28]. Thirdly, $\tau(X^0)$ is always less than that predicted in the strong confinement limit pointing to the presence of electron-hole correlations. In a two level model, the recombination lifetime is given by[29]:

$$\frac{1}{\tau} = \frac{2n\pi e^2 f_{osc}}{3\lambda_{PL}^2 \varepsilon_0 c m_0} \qquad (2)$$

where n is the refractive index, 3.5 for GaAs, $\lambda_{PL}$ is the emission wavelength and $f_{osc}$ the transition oscillator strength. In the strong confinement limit the minimum lifetime corresponds to the maximum overlap, $|\langle \Psi_e | \Psi_h \rangle|^2 = 1$. In this limit, an emission energy of 1.3 eV leads to an oscillator strength of 9.88 and a lifetime of 1.15 ns, significantly higher than the measured $X^0$ lifetime at 1.3 eV, ~ 0.8 ns. Finally, equations 1 and 2 predict a slight dependence of the recombination lifetime on dot energy. Figure 3 includes the calculated lifetime from Eqs. 1 and 2 as a function of emission energy, once again assuming maximum overlap. The change in measured lifetime with increasing emission energy is larger than that predicted by the strong confinement model. This shows that the strong confinement model becomes more and more inappropriate as the PL energy increases, presumably a result of a softening of the dot confinement.

**Effect of Dot Charge on Recombination Lifetime and Dot energies**

The discussion above shows that the radiative lifetime is sensitive to the nature of the exciton confinement. Our results in Fig.s 3 and 4 show that charging an exciton with an electron produces a small effect whereas charging with a hole produces a much larger effect. The natural interpretation is that the electron is close to the strong confinement limit but that the hole is in an intermediate regime. Our results suggest that in the $2X^0$ or $X^{1+}$ states, the Coulomb repulsion between the two holes causes the hole wave function to expand laterally such that the overlap with the electron wave function is decreased. Within a configuration picture, this is consistent with significant Coulomb-induced hybridization of the hole s-like orbital with the p-like (and possibly also d-like) orbitals which results in a net decrease in the overlap with the electron s-like orbital.

In addition to the lifetimes, we can use in addition the PL energies and charging voltages to deduce the effects of charging on the electron and hole wave functions. Again, we present robust conclusions by analyzing data from many dots. We describe the charging of the dots with a simple phenomenological Coulomb blockade (CB) model. The model simply re-casts the experimental data, gate voltage extents and PL energy shifts, in terms of Coulomb energies.

Each Coulomb energy is defined as $E_{ab}^{\alpha\beta}$, where ab identifies the type of Coulomb interaction (ee for electron-electron, hh for hole-hole and eh for electron-hole), α identifies the number of electrons in the dot, and β identifies the number of holes. Table 1a lists the energies of the exciton configurations $X^0$, $2X^0$, $X^{1-}$ and $X^{1+}$; the "no hole" states e and 2e; and the "no electron" states h and 2h in terms of the Coulomb energies, the electrostatic potential $e(V_0-V_g)\lambda^{-1}$, the single particle energy gap $E_0$, the interaction between a charge in the dot with its image charge in the back contact $E_i$, and the single electron ionization energy $E_c$. $V_g$ is the gate voltage, $V_0$ the Schottky barrier height (0.62 V) and $\lambda$ is the device lever arm (6.45 for sample A and 7.00 for sample B) which converts the applied bias into potential energy at the location of the dot. $E_i$ is -1.1 meV for both samples[17]. This parameterization of the experiment does not assume strong confinement, as was the case previously[12, 17, 22, 23, 30], or a particular form of the confinement potential[22, 23].

Experimentally, it is the difference in PL energies on charging and the $V_g$-extents of the charging plateau which contain information on the Coulomb energies. Each PL energy depends on the energy difference between the initial and final states, giving the results in Table 1b. To determine the voltage extent of each charging plateau, the biases at which the dot charges with a single electron are determined. For example, $X^0$ turns on when the single hole state h and the $X^0$ exciton state are degenerate and turns off when the $X^0$ and $X^{1-}$ states are degenerate, Fig. 1. The gate voltage extents of $X^{1+}$ and $X^0$, labeled $\Delta V(X^{1+})\lambda^{-1}$ and $\Delta V(X^0)\lambda^{-1}$, are not the same owing to the different Coulomb terms involved in each case, Table 1c. Charging $2X^0$ and $X^{1-}$ involves occupying the electron p orbital, and therefore introduces yet more CB parameters without yielding any further insights into the CB parameters involving the s orbitals. However, the final state of the $X^{1-}$ contains a single electron and this enables us to deduce CB parameters involving single electron charging in the absence of a hole.

Over the voltage extent of $X^{1-}$ the stable no-hole state changes from the vacuum, to a single electron to 2 electrons[17, 30]. Consequently, at the low bias side of $X^{1-}$ the electron in the final state after recombination tunnels on a timescale of ~10 ps out of the dot. Similarly, at the high bias side the final state is also unstable with respect to tunneling, in this case an electron from the back contact tunnels into the dot after recombination. In the regions of rapid final state tunneling, the $X^{1-}$ PL is significantly broadened, and close to the final state degeneracies, there is a blue-shift in the emission wavelength as a consequence of a more coherent hybridization[31]. Fig. 6a shows an example of the measured PL peak energy of $X^{1-}$ as a function of bias from a single dot from sample B. As the bias increases, the peak energy shifts to the blue through to the Stark effect[32]. The Stark shift is fitted according to $E_{PL} = E_0 - pF + \beta F^2$ with parameters $E_0 = 1.298$ eV, $p/e = 2.4$ nm and $\beta = 0.6$ μeV/(kV/cm)$^2$. In addition to the Stark shift there are two distinct blue shifts in the PL energy at -0.35 V and -0.22 V, highlighted in Fig. 6b by subtracting the Stark shift fit from the data. These two blue shifts correspond directly to the changing final state of $X^{1-}$ and provide a direct experimental measurement of the two voltages. Consequently this technique provides a direct measurement of $E_{ee}^{20}$, Table 1. It should be noted that this blue shift of the PL depends on the tunneling time which tends to decrease as the PL energy decreases. For dots on the red end of the ensemble, the hybridization is difficult to resolve. We have determined $E_{ee}^{20}$ in this way for about half of the dots with an uncertainty of about 1.5 meV.

The expressions for $\Delta(0e)\lambda^{-1}$ and $E_{PL}(X^0) - E_{PL}(X^{1-})$ in Table 1 are identical, providing a simple consistency check. Fig. 7a compares, for each dot, $E_{PL}(X^0) - E_{PL}(X^{1-})$ with $\Delta(0e)\lambda^{-1}$. Within experimental error there is good agreement between the two parameters, highlighted in Fig. 7b where the difference between $E_{PL}(X^0) - E_{PL}(X^{1-})$ and $\Delta(0e)\lambda^{-1}$, which should be zero, is plotted as a function of dot PL energy. There is no dependence on $X^0$ PL energy and we find an average value of 0.289 ± 0.916 meV. This result provides strong support for the reliability of our model, in particular the interpretation of the $X^{1-}$ PL and the use of the lever arm to convert applied bias to electrostatic potential.

It is now possible to plot the initial and final state energies from Table 1a by using the measured gate voltage extents of the exciton plateaus and the energy differences between the PL lines along with the above method to determine $E_{ee}^{20}$. Results for one particular dot are shown in Fig. 1. We now attempt to determine the individual Coulomb energies, not just sums and differences. For $X^0$ and $X^{1-}$, there are four Coulomb terms, $E_{ee}^{20}, E_{ee}^{21}, E_{eh}^{11}$ and $E_{eh}^{21}$. There are four measured quantities, $\Delta(0e)\lambda^{-1}, \Delta(1e)\lambda^{-1}, \Delta(X^0)\lambda^{-1}$ and $E_{PL}(X^0) - E_{PL}(X^{1-})$. However, the equality of $\Delta(0e)\lambda^{-1}$ with $E_{PL}(X^0) - E_{PL}(X^{1-})$ removes one independent expression from Tables 1b and 1c preventing a complete determination of the Coulomb energies. We obtain approximate results by assuming that $E_{ee}^{20} = E_{ee}^{21} (= E_{ee}^{22})$, i.e. that the electron-electron Coulomb energy is independent of the hole occupation, equivalently that the electron wave function is frozen. This assumption is motivated by the small change in the radiative lifetime on electron charging and existing understanding of the biexciton lifetime[7]. We label the fixed electron-electron Coulomb term $E_{ee}$ and determine $E_{ee}$, $E_{eh}^{11}$, and $E_{eh}^{21}$ for all dots where there is a measurable hybridization-induced blue shift in the $X^{1-}$ PL, Figs. 7c and 7d. The important result is that $E_{eh}^{21}$ is found to be, on average, 1.78 ± 1.98 meV smaller than $E_{eh}^{11}$, Figs. 7d and 7e. In other words, charging the exciton with an extra electron changes the hole wave function leading to a decrease in the electron-hole Coulomb energy. The change with charge is 5-10% for the energies compared to 25% for the lifetimes showing that the energies are less sensitive to Coulomb correlations than the decay rates, in agreement with existing theoretical work[27].

Turning to $X^{1+}$ and $2X^0$, and retaining the frozen electron approximation, the introduction of a second hole introduces 5 new Coulomb terms $E_{hh}^{02}, E_{hh}^{12}, E_{hh}^{22}, E_{eh}^{12}, E_{eh}^{22}$ yet there are only three new experimental parameters, the voltage extent of $X^{1+}$ and the 2 new energy differences between the PL lines, Table 1. We are therefore unable to determine $E_{eh}^{12}$. However, Table 1 shows that the hole-hole Coulomb energy without electrons, $E_{hh}^{02}$, can be determined from the experiment through:
$E_{hh}^{02} = \Delta V(X^{1+})\lambda^{-1} + [E_{PL}(X^{1+}) - E_{PL}(2X^0)] - [E_{PL}(X^0) - E_{PL}(X^{1+})] + 2E_i$,
a result which also holds without the frozen electron wavefunction approximation. Unfortunately, only a few dots show clear $2X^0$ PL emission and a well defined $X^{1+}$ extent. For the dot shown in Fig. 1, $E_{hh}^{02}$ is determined to be 30.4 ± 1.5 meV, which compares to 29.6 ± 1.5 meV for $E_{ee}^{20}$ for the same dot. Similar values are obtained from a further two dots from the entire data set. The conclusion is that $E_{hh}^{02}$ is surprisingly small. The hole effective mass is significantly larger than the electron effective mass[22] leading to more localized single particle hole wave functions than electron wave functions. In the strong confinement limit, this would lead to much larger, perhaps ~50% larger, hole-hole Coulomb energies than electron-electron Coulomb energies. This is clearly not the case: the holes are not in the strong confinement limit. Recent theoretical calculations using pseudopotential atomistic wavefunctions agree well with our experimental results and extend the principle of an intermediate hole confinement regime to explain many body effects in highly charged excitons[33]. We expect, although are unable to confirm, that $E_{hh}^{12}$ and $E_{hh}^{22}$ are within a few meV of $E_{hh}^{02}$ as the results point to large changes in hole wave functions on hole charging but relatively small changes on electron charging.

## Conclusions

We present statistics on the recombination lifetimes and energies of the neutral exciton, the neutral biexciton, the positively charged exciton and negatively charged exciton in single self-assembled InGaAs quantum dots. There are significant dot-to-dot fluctuations in all these parameters for a particular neutral exciton emission energy. Nevertheless, in relation to our data set, some definite conclusions can be reached:

1. $\tau(X^{1+})$ is always larger than $\tau(X^0)$. Averaged over our data, $\tau(X^{1+}) / \tau(X^0) = 1.58 \pm 0.55$.
2. $\tau(X^{1-})$ is for 94% of the dots larger than $\tau(X^0)$. Averaged over our data, $\tau(X^{1-}) / \tau(X^0) = 1.25 \pm 0.18$.
3. $2\tau(2X^0)$ is always larger than $\tau(X^0)$. Averaged over our data, $2\tau(2X^0) / \tau(X^0) = 1.30 \pm 0.2$.
4. $E_{PL}(X^{1-})$ is always smaller than $E_{PL}(X^0)$. Averaged over our data, $E_{PL}(X^{1-}) - E_{PL}(X^0) = -5.3 \pm 0.42$ meV.
5. $E_{PL}(2X^0)$ is always larger than $E_{PL}(X^{1-})$ but less than $E_{PL}(X^0)$. Averaged over our data, $E_{PL}(2X^0) - E_{PL}(X^0) = -2.2 \pm 0.86$ meV.
6. $E_{PL}(X^{1+})$ is, in 74% of our data, blue-shifted with respect to $E_{PL}(X^0)$. Averaged over our data, $E_{PL}(X^{1+}) - E_{PL}(X^0) = 0.85 \pm 1.35$ meV.

By converting the voltage plateau into energies using the lever arm model, we can parameterize these results in terms of Coulomb energies. We find that:

1. The electron-hole Coulomb energy is larger than the electron-electron Coulomb energy, typically by $5.3 \pm 0.4$ meV, for all dots.
2. The electron-electron and hole-hole Coulomb energies are the same to within our experimental error of a few meV.
3. The electron-hole Coulomb energy of an electron-hole pair is reduced by about 2 meV on charging the dot with a further electron.

To within the fidelity of the experiment, limited in some cases by uncertainties in determining the single-electron Coulomb blockade voltages and more generally by the dot-to-dot fluctuations, we can state that these results are consistent with a strong confinement model for the electrons in which the electron single particle wave function is largely unchanged on charging and consistent with an intermediate confinement model for the holes in which the single particle hole wave function extends laterally slightly in the presence of an electron and significantly in the presence of an additional hole.

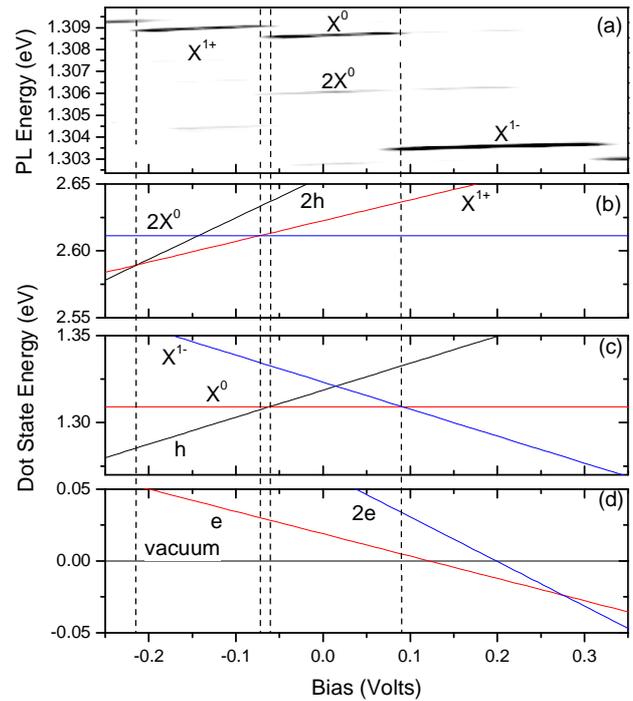

Figure 1 : (Colour online) (a) Grey scale plot showing the charging behavior as a function of bias of the time-integrated PL from a single dot from sample A at 4.2 K. Data were taken with a pulsed (20 MHz) 826 nm laser delivering 20 nW / μm² excitation power at the sample. White corresponds to 200 counts, black to 1500 counts. The neutral exciton, $X^0$, negatively charged exciton, $X^{1-}$, positively charged exciton, $X^{1+}$, and biexciton, $2X^0$, are labeled. (b) and (c) show the exciton energies of the initial states of the dot shown in (a) as determined by the Coulomb blockade model summarized in Table 1. (d) shows the energies of dot states containing no holes: the vacuum state is the final state of the $X^0$ transition; state e is the final state of the $X^{1-}$ transition.

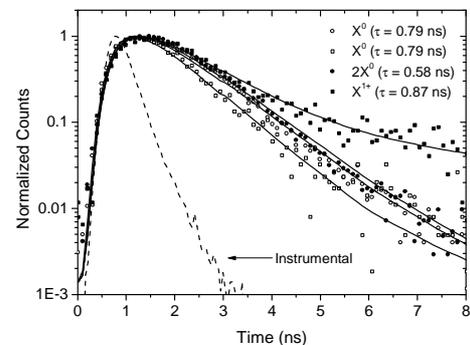

Figure 2 : Normalized radiative lifetime data from $X^0$, $X^{1-}$, $X^{1+}$, and $2X^0$ for the dot shown in Fig 1a. Each decay is taken under the same excitation conditions as Fig 1a and from the centre of each exciton's voltage plateau. Convolution fits to the data are shown, along with the instrumental response of the system (FWHM of ~400ps). The fitted recombination lifetimes are 0.79 ns, 0.84 ns, 0.87 ns and 0.58 ns for $X^0$, $X^{1-}$, $X^{1+}$, and $2X^0$,

respectively. The $X^{1+}$ decay is biexponential with a secondary lifetime of 6.7 ns, a process attributed to hole recapture from the capping layer/blocking barrier interface. All other decays are single exponentials. The integration time to record each decay is 300 seconds.

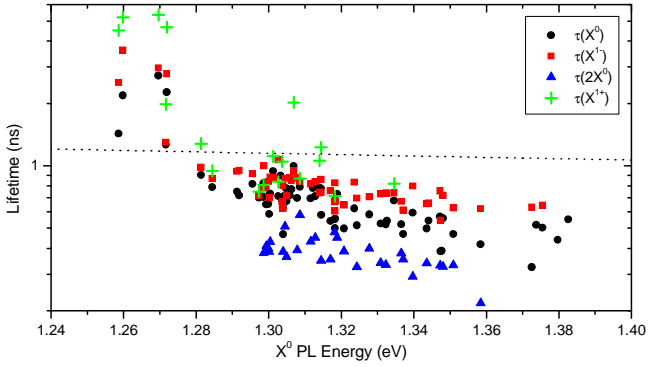

Figure 3 : (Colour online) The measured radiative lifetime for $X^0$, $X^{1-}$, $X^{1+}$, and $2X^0$ from almost 80 dots from both sample A and sample B as a function of neutral exciton PL energy. The dashed line is the calculated lifetime in the strong confinement limit with maximum wavefunction overlap.

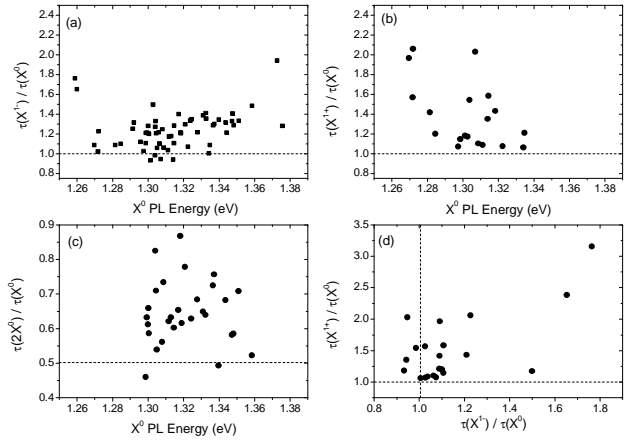

Figure 4 : (a) $\tau(X^{1-})/\tau(X^0)$, (b) $\tau(X^{1+})/\tau(X^0)$, and (c) $\tau(2X^0)/\tau(X^0)$ plotted against the $X^0$ PL energy. Each point represents the result from one particular quantum dot. (d) shows $\tau(X^{1-})/\tau(X^0)$ versus $\tau(X^{1-})/\tau(X^0)$ for the subset of data containing a reliable $X^{1+}$ radiative decay time.

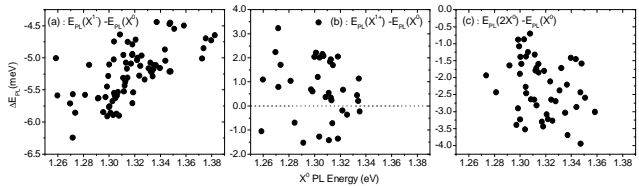

Figure 5 : The PL energy differences (a) $X^{1-}$ with respect to $X^0$, (b) $X^{1+}$ with respect to $X^0$ and (c) $2X^0$ with respect to $X^0$.

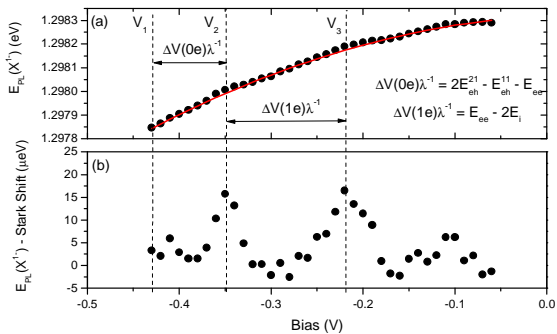

Figure 6 : (a) The measured $X^{1-}$ PL peak energy over the bias extent of the $X^{1-}$ plateau from a single dot from sample B ($\lambda = 7$). $V_1$ represents the bias voltage at which $X^{1-}$ is charged from $X^0$. The overall bias dependence is dominated by the Stark shift, highlighted by the solid line with fit parameters $E_0 = 1.298$ eV, $p/e = 2.4$ nm and $\beta = 0.6$ $\mu$eV/(kV/cm)$^2$. The blue shifts in the PL energy at $V_2$ and $V_3$ correspond to electron tunneling in the exciton final state. (b) The effect of the final state tunneling on the emission energy is highlighted through subtracting the Stark shift fit from the original data.

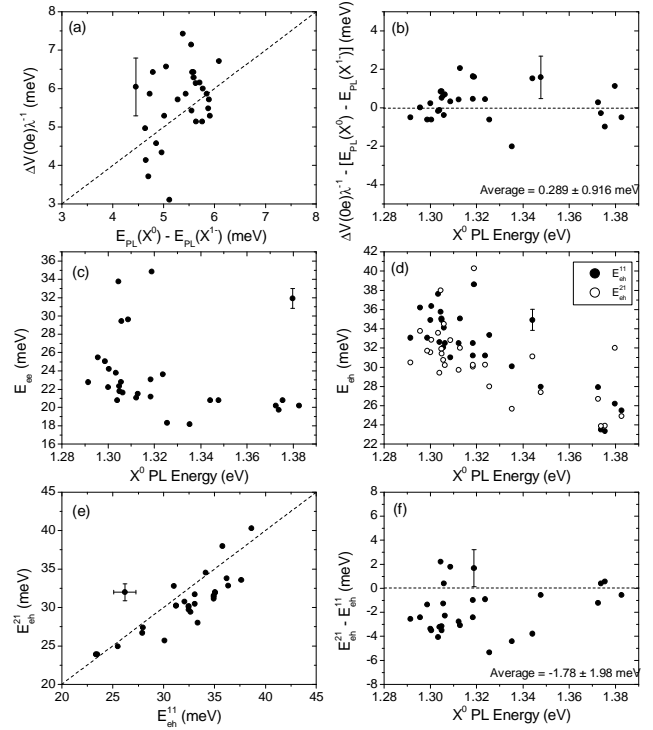

Figure 7 : (a) $e(V_2 - V_1)\lambda^{-1}$ versus $E_{PL}(X^0) - E_{PL}(X^{1-})$. Each point represents the measurement on one particular dot; the dashed line $e(V_2 - V_1)\lambda^{-1} = E_{PL}(X^0) - E_{PL}(X^{1-})$. (b) $e(V_2 - V_1)\lambda^{-1} - E_{PL}(X^0) - E_{PL}(X^{1-})$ plotted against $X^0$ PL energy. (c) Electron-electron and (d) electron-hole $X^0$ and $X^{1-}$ Coulomb energies versus $X^0$ PL energy. (e) shows the electron-hole Coulomb energies for $X^{1-}$, $E_{eh}^{21}$, plotted against the electron-hole Coulomb energy for $X^0$, $E_{eh}^{11}$. (f) $E_{eh}^{21} - E_{eh}^{11}$ versus $X^0$ PL energy. One point in each figure shows a representative error bar.

| (a) State of QD | Initial State Energy |
|---|---|
| $E(e)$ | $(V_0 - V_g)\lambda^{-1} - E_c - E_i$ |
| $E(2e)$ | $2(V_0 - V_g)\lambda^{-1} - 2E_c + E_{ee}^{20} - 4E_i$ |
| $E(h)$ | $E_0 - (V_0 - V_g)\lambda^{-1} + E_c - E_i$ |
| $E(2h)$ | $2E_0 - 2(V_0 - V_g)\lambda^{-1} + 2E_c + E_{hh}^{02} - 4E_i$ |
| $E(X^0)$ | $E_0 - E_{eh}^{11}$ |
| $E(X^{1-})$ | $E_0 + (V_0 - V_g)\lambda^{-1} - E_c - 2E_{eh}^{21} + E_{ee}^{21} - E_i$ |
| $E(X^{1+})$ | $2E_0 - (V_0 - V_g)\lambda^{-1} + E_c + E_{hh}^{12} - 2E_{eh}^{12} - E_i$ |
| $E(2X^0)$ | $2E_0 + E_{hh}^{22} + E_{ee}^{22} - 4E_{eh}^{22}$ |
| (b) Energy Difference | Energy |
| $E_{PL}(X^0) - E_{PL}(X^{1-})$ | $2E_{eh}^{21} - E_{eh}^{11} - E_{ee}^{21}$ |
| $E_{PL}(X^{1+}) - E_{PL}(2X^0)$ | $E_{hh}^{12} - E_{hh}^{22} + 4E_{eh}^{22} - 2E_{eh}^{12} - E_{eh}^{11} - E_{ee}^{22}$ |
| $E_{PL}(X^0) - E_{PL}(X^{1+})$ | $2E_{eh}^{12} - E_{eh}^{11} - E_{hh}^{12}$ |
| (c) Plateau Extent | Energy |
| $\Delta V(X^0)\lambda^{-1}$ | $2E_{eh}^{11} - 2E_{eh}^{21} + E_{ee} - 2E_i$ |
| $\Delta V(0e)\lambda^{-1}$ | $2E_{eh}^{21} - E_{eh}^{11} - E_{ee}^{21}$ |
| $\Delta V(1e)\lambda^{-1}$ | $E_{ee}^{20} - 2E_i$ |
| $\Delta V(X^{1+})\lambda^{-1}$ | $E_{hh}^{02} + E_{hh}^{22} - 2E_{hh}^{12} + 4E_{eh}^{12} - 4E_{eh}^{22} + E_{ee}^{22} - 2E_i$ |

Table 1 : (a) Expressions describing the energies of all states with occupied s orbitals. The Coulomb interactions are parameterized by labelling the Coulomb energies $E_{ab}^{\alpha\beta}$, where ab identifies the type of Coulomb interaction (ee for electron-electron, hh for hole-hole and eh for electron-hole), $\alpha$ identifies the number of electrons in the dot and $\beta$ identifies the number of holes. (b) The PL energy differences and (b) the Coulomb blockade gate voltage extents for the various excitons.